# Large dynamic range SPR measurements in the visible using a ZnSe prism

John Canning,[1] Jinh Qian,[1] Kevin Cook[1]

[1] interdisciplinary Photonics Laboratories, School of Chemistry, The University of Sydney, NSW 2006, Australia
*Corresponding author: john.canning@sydney.edu.au

Large dynamic index measurement range ($n$ = 1 to $n$ = 1.7) using surface plasmon resonance (SPR) shifts is demonstrated with a ZnSe prism at 632.8 nm, limited by the available high index liquid hosts. In contrast to borosilicate based SPR measurements where angular limitations restrict solvent use to water and requires considerable care dealing with Fresnel reflections, the ZnSe approach allows SPR spectroscopies to be applied to a varied range of solvents. An uncertainty in angular resolution between 1.5 and 6°, depending on the solvent and SPR angle, was estimated. The refractive index change for a given glucose concentration in water was measured to be $n = (0.114 \pm 0.007) /\%[C_6H_{12}O_6]$. Given the transmission properties of ZnSe the processes can be readily extended into the mid infrared. © 2013 Optical Society of America

Surface plasmon resonances (SPR), or surface waves, in metals stretch back some time, perhaps to Wood's observations of anomalous reductions in scattered light within metal gratings [1]. They are seeing new light in recent years because of the high sensitivity of these resonances to changes in the local environment of a conductive surface and have been particularly exciting for biosensing interface applications since they are sensitive and label free [2-4]. Typically, because of the large photon-SPR wave mismatch, evanescent field coupling of reflected *p*–polarised light above the cut-off for total internal reflection at a dielectric/metal interface is required to excite the surface plasmon resonance on the other side of a metal film. Prisms are used in commercial instruments to avoid air-dielectric challenges of getting light into the prism to launch at the SPR angle, $\theta_{SPR}$. Alternative methods include end coupling using various wave guiding configurations [5,6], grating based diffraction assisted angular coupling [7-9], optical waveguide coupling [10], multimode optical fibre waveguide coupling [6], single-mode optical fibre waveguide and grating coupling combinations [11-13], scattering [14] and many more.

Evanescent field coupling is generally most often done with either triangular or circular prisms configured in the straight-forward so-called Kretschmann setup [15], shown in Figure 1. These prisms are typically borosilicate, silica or acrylate based; for water, these materials are close to the practical coupling edge and in the case of silica, SPR excitation is not possible at 632.8 nm, a commonly used HeNe laser wavelength. Hence, commercial SPR instruments tend to use higher index borosilicate glass for transparency and reduced scattering in the visible. From the required wave vector matching condition, the SPR angle at particular wavelength, $\theta(\lambda)_{SPR}$, can be related to the prism index, $n(\lambda)_p$, as:

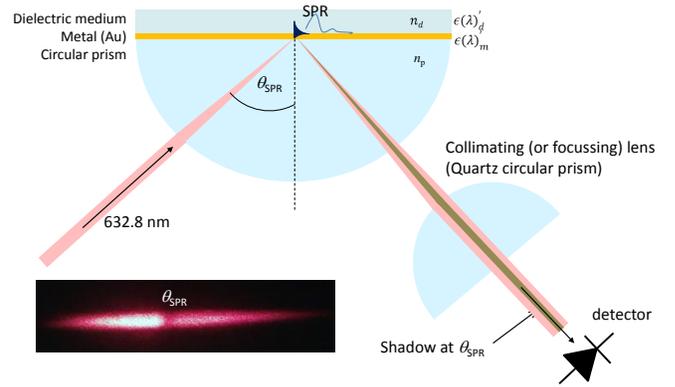

$$\theta(\lambda)_{SPR} = sin^{-1}\left[\frac{1}{n(\lambda)_p}\left(\frac{\epsilon(\lambda)'_d \epsilon(\lambda)'_m}{\epsilon(\lambda)'_d + \epsilon(\lambda)'_m}\right)^{\frac{1}{2}}\right] \quad (1)$$

where $\epsilon(\lambda)'_d$ is the real part of the dielectric constant of the dielectric above the metal, $\epsilon(\lambda)'_m$ is the dielectric constant of the metal. For the common situation involving a Au film on borosilicate glass prism ($n_p \sim 1.515$ @632.8 nm), the calculated SPR coupling angle at 632.8 nm will be $\theta_{SPR} \sim 44°$ and $\theta_{SPR} \sim 74°$ when air and water are the respective dielectric media on top of the metal. Water is most often used for biodiagnostics with SPR in commercial instruments but already the dynamic range can be quite limited given that the SPR angle is close to practical limits involved with probing angular space, usually $\theta < 80°$ [16]. Such low angles from the surface also create other optical reflection, refraction and dispersion challenges that can necessitate a more complex optical configuration. Typically, with water a working refractive index range of 1.33 to 1.47 is possible. For biosensing applications the angle is fixed and the SPR shift monitored over a small range as the intensity varies when the local index changes arising from the interfacial interaction being monitored. In such cases the system can have very high resolution but the range is limited to the steepness of the SPR wall. Sometimes, the fixed wavelength measurement is combined with the tracking of the SPR resonance shift to utilise the full range. Unfortunately, not all chemical or biosensing requirements work within this range or with pure water as a host medium. Many solvent hosts have higher refractive indices which tend to lie outside the practical range. This point is emphasised in Figure 2, a calculation of the SPR angle as a function of prism index for a range of dielectric solvents. For pure silica, only in air can an SPR be excited because the refractive index is too low. This is improved with borosilicate where an SPR can be excited in both air and water but not much else that has an index above 1.4. Further, depending on how the measurements are conducted, SPR sensors face a challenge given some evidence that water, a polar solvent, can interact with a Au surface to generate different properties at the interface than at the bulk and therefore different outcomes between events [17], accounting for some the experimental variation observed in the literature. The ability to use higher index, non-polar solvents is therefore desirable.

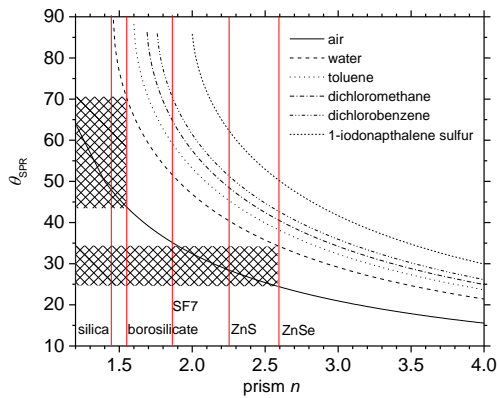

**Fig 2.** The calculated SPR angle, $\theta_{SPR}$, as a function of prism index for various dielectric solvents on Au. The specific refractive index for silica, borosilicate, SF7, ZnS and ZnSe are shown. Only the latter two have $\theta_{SPR}$ for all the solvents with ZnSe offering the greatest tuning range.

Figure 2 illustrates that the simplest approach to overcome the challenges of higher index dielectric media and the need for greater dynamic range is to raise the prism index. Practically, this allows SPR to be excited as well as reducing the required $\theta_{SPR}$ relative to the borosilicate. Reducing the SPR angle also allows a greater angular span to be achieved and therefore increase the dynamic range of the system to allow more solvent hosts to be used; this is illustrated by ZnS and ZnSe both of which are able to measure all the example solvents in Figure 2. The higher index greatly reduces the dispersion in $\theta_{SPR}$ between air and water ($10^0$ with ZnSe; $28^0$ with borosilicate) which potentially generates other benefits. Higher index SF7 prisms ($n \sim 1.7$ [18]) have previously been used to reduce the SPR angle for water to $\theta_{SPR} \sim 57$ [19]. Other higher index materials have also been used particularly for near and mid-infrared SPR work given many, such as $CaF_2$ (only slightly higher at $n = 1.4339$),

ZnS and ZnSe, transmit in this region and SPR resonances are much narrower. For example, $CO_2$ detection using a mid-infrared SPR sensor and a $CaF_2$ prism has been demonstrated [19] and long range SPR assisted Goos–Hänchen and Imbert–Fedorov shifts in ZnSe prism at 1550 nm have been reported [20], whilst ZnSe prism was also used in the infrared to monitor phospholipid membrane growth at 632.8 nm, near infrared ~ (1.4-1.5) μm and at 10.6 μm [21]. In other non-planar formats, ZnSe is often used; for example, as a composite nanoparticle with Au helped to enhance photoluminescence via SPR excitation [22].

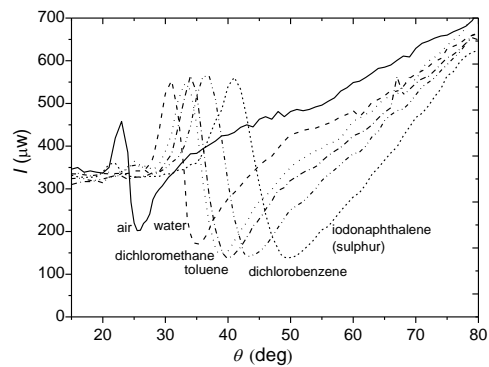

**Fig 3.** The experimentally measured SPR profiles for a Au film on ZnSe prism with different dielectric solvents. The graphs are raw data and not corrected for variations in Fresnel reflection coefficients. Chirping, or dispersive broadening, of the signal with higher angle is observed.

It stands to reason that a high index substrate such as ZnSe can transmit visible light should outperform the borosilicate standard. ZnSe is in a class of rigid polar crystal structures capable of supporting surface phonon polaritons, which can potentially be exploited for metamaterial applications [23], although these are not considered here. In this work, we use ZnSe as a dielectric prism because it has the largest potential tuning range whilst retaining sufficient visible transmission at common wavelengths, such as 632.8 nm, for visible SPR and has a sufficiently high index ($n_p = 2.591$ @ 632.8 nm) to shift $\theta_{SPR}$ at these wavelengths by a significant margin. Much existing biosensing applications carried out in water with borosilicate glass prisms can be undertaken by simply substituting the prism and adjusting for the change in angle. From equation (1), and summarised in Table 1, $\theta_{SPR}$ = 26.5° and 37.6°, using ZnSe for air and water respectively, significantly lower than that of borosilicate. Also indicated in Figure 2, a ZnSe prism greatly expands the potential solvent access to unprecedented levels. For example, it should still be possible to excite an SPR even when the dielectric refractive index is as high as 1.7.

Using the configuration illustrated in Figure 1, where a circular ZnSe prism is used, the SPR profile is measured experimentally. The Au layer of 45 nm was deposited directly onto the ZnSe prism by sputtering – this avoids the need for using ZnSe slides and also reduces possible etalon and wave guiding effects between prism and slide (which are not uncommon in these setups suggesting slight variations in index between manufacturers). Because of the high refractive index, prism curvature and a finite Gaussian laser beam width of 0.8 mm, there is sufficient angular dispersion that leads to a portion of the centre light meeting the SPR condition and the rest on either side of $\theta_{SPR}$, experiencing predominantly Fresnel reflections. The inset of Figure 1 is an image of the reflected light showing the dark middle region where light is absorbed as SPR. In the configuration, a quartz circular prism is used as a lens to focus this image onto a power meter. In this way the average power across the image is measured, greatly simplifying the experimental configuration during rotation but reducing the SPR signal contrast. Figure 3 shows the SPR profile for all the dielectric media listed in Figure 2. Table 1 summarises all the relevant data for comparison.

maximum just beyond the SPR region. The choice of method is affected by the quality of results. The results for ZnSe and each particular solvent are also shown in Table 1. Consistent with the observed angular dispersion, which causes the significant broadening in the SPR profile, the effective resolution degrades with increasing solvent index.

|  | Air | Water $H_2O$ | Dichloromethane $CH_2Cl_2$ | Toluene $C_7H_8$ | Dichlorobenzene $C_6H_{14}Cl_2$ | 1-iodonaphthalene (sulphur) $C_{10}H_7I(S)$ |
|---|---|---|---|---|---|---|
| n | 1 | 1.332 | 1.4242 | 1.4936 | 1.5241 | 1.7 |
| $\theta_{SPR}$ silica (n = 1.45) | 48°  46° | - | - | - | - | - |
| $\theta_{SPR}$ borosilicate (n = 1.515) | 44°  44° | 70.4°  70° | - | - | - | - |
| $\theta_{SPR}$ ZnS (n = 2.35) | 28.5°  - | 40.5°  - | 48.5°  - | 45.4°  - | 51.4°  - | 62.4°  - |
| $\theta_{SPR}$ ZnSe (n = 2.591) | 24.5°  25.7°  (1.6°) | 34.3°  35.1°  (2.5°) | 38.4°  38.9°  (3.9°) | 40.4°  39.7°  (4.6°) | 43.0°  43.3°  (5.4°) | 50.2°  49.6°  (5.9°) |

**Table 1.** Summary of results showing refractive indices of materials and both calculated and measured (below) SPR angles for different solvents and prism materials. For ZnSe the estimated experimental resolution uncertainty, $\Delta\theta_{SPR}$, is shown bottom in parenthesis.

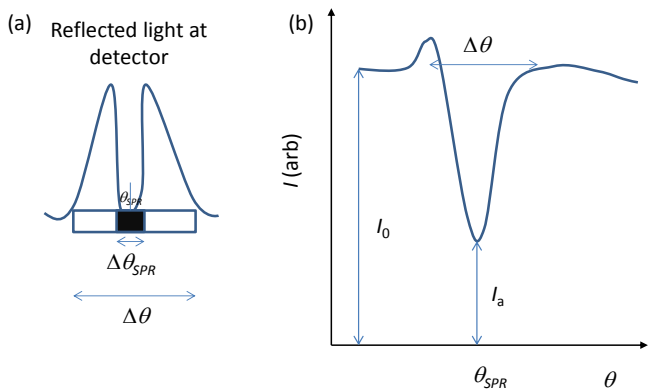

**Fig 4.** Idealised schematic of (a) the measured profile at the detector over a finite angle dispersion, $\Delta\theta$, and (b) the measured profile as a function of angle. $\Delta\theta$ determines the effective resolution of $\theta_{SPR}$, $\Delta\theta_{SPR}$, which can be related back to the measured light outside of the SPR ($I_a$) for the case of a uniform film with perfect $\theta_{SPR}$ signal contrast where $I_{\theta(SPR)} = 0$.

The angular spectra do not go to zero primarily because the power is an average measurement around $\theta_{SPR}$; the contrast and the profile resolution may be improved by employing an aperture to reduce the observed angle and cut out the reflected light whilst rotating the prism and power meter. This was not done here to keep the experimental setup as simple as possible. However, assuming an ideal uniform layer of Au on the ZnSe prism, such that the ideal SPR profile reaches zero, one can define an effective angular resolution uncertainty, $\Delta\theta_{SPR}$, of the experimental setup from the background light component directly from the measured spectra noting the following:

$$\Delta\theta_{SPR} \sim \frac{1}{2}\left(\frac{I_0 - I_a}{I_0}\right)\Delta\theta \quad (2)$$

where $I_0$ is the intensity of the light detected outside the SPR region, $I_a$ is the light intensity transmitted and measured at $\theta_{SPR}$ and $\Delta\theta$ is the angular space over which the measurement is taken which can be directly derived from the experimental image (inset Figure 1) or be approximately estimated from the angular intensity profile where the intensity on either side of the SPR equals a

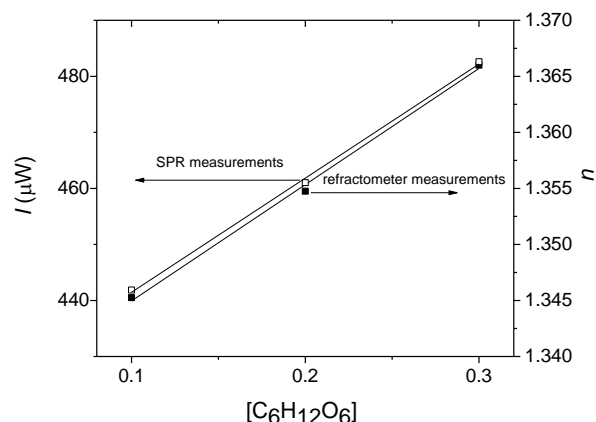

**Fig 5.** Change in refractive index when glucose (10, 20 and 30 %) is added to distilled, deionised water measured by fixed angle SPR at ZnSe/Au interface as optical intensity changes and also measured directly on a wavelength refractometer at 632.8 nm.

In many experiments, the change in refractive index as a function of concentration is small and therefore not measured as a function of resonance shift but rather an intensity change in reflected light using a fixed angle excitation of the SPR. The change in water refractive index with added glucose was measured using a fixed angle excitation on the shorter wavelength side of the Au SPR through the ZnSe prism and subsequent changes in optical intensity (which increases with increasing refractive index), shown in Figure 5. The results are compared with direct refractometer measurements and show 1:1 correspondence on the linear part of the SPR resonance. A variation in (0.114 ± 0.007) refractive index units /% fraction of glucose (r.i./%) is obtained. With ZnSe, unlike conventional SPR measurements using borosilicate prisms, from Figures 2 and 3, other species can be similarly characterised in different solvent hosts.

Conclusions

Using a high index ZnSe prism very wide dynamic range of an SPR setup measuring refractive indices from 1 to 1.7 has been demonstrated with available solvent hosts. Unlike typical setups using borosilicate for water based measurements where angular SPR position starts close to $70^0$, with not much more than $10^0$ tuning range up to $80^0$ greatly restricting solvent access with refractive index above water, this setup is able to excite SPR from a low of $\sim (34\text{-}35)^0$ in Au providing more than $45^0$ of potential tuning. Almost any arbitrary solvent of higher index can be used greatly extending SPR based spectroscopies beyond water based measurements. As the prism index increases, the SPR angle is reduced and, notably, so too is the angular dispersion between solvent indices. Borosilicate is limited both for physical and practical reasons to water or solvents not much above in index; similarly, silica cannot excite SPR via evanescent field from reflected light above the critical angle with water on Au. Further, undertaking measurements in water using ZnSe, at a more accessible SPR angle well below that of borosilicate glass, greatly relaxes some of the alignment and additional photonic challenges raised by tight angles. A simple method of determining a qualitative experimental uncertainty in the SPR angle was proposed based on the difficulty of avoiding light capture outside of the SPR angle. Measurements for glucose in water were also carried out using fixed angle interrogation. It is noted that the increased dynamic range should not be limited to the visible and will have potential benefits in the mid infrared where the SPR resonances are expected to be narrower. These concepts can be extended to consider other plasmon excitation routes within high index dielectric material, including waveguide and grating based.

*Acknowledgements*: This work was funded by the Australian Research Council grants DP DP140100975 and FT110100116. J. Qian acknowledges an *i*PL Summer Scholarship.